\documentclass[conference]{IEEEtran}
\IEEEoverridecommandlockouts

\usepackage{amsmath,amssymb,amsfonts}
\usepackage{mathtools}

\usepackage{algorithmic}
\usepackage{graphicx}
\usepackage{textcomp}
\usepackage{float}
\usepackage{adjustbox}
\usepackage{cite}

\usepackage{comment}

\usepackage[dvipsnames]{xcolor}

\usepackage{tikz}
\usetikzlibrary{shapes.geometric, arrows.meta, positioning, fit, backgrounds, calc, patterns, shadows}

\usepackage{pgfplots}
\pgfplotsset{compat=1.18}
\usepgfplotslibrary{groupplots}

\usepackage{hyperref}

\pgfplotsset{
  tiny,
  legend style={
    at={(0.03,0.03)},
    anchor=south west,
    font=\footnotesize,
    draw=none,
    fill=white,
    fill opacity=0.8,
    text opacity=1,
  },
  every axis plot/.append style={
    line width=1.2pt,
    solid,
  },
}

\definecolor{gold}{RGB}{212,175,55}
\definecolor{navy}{RGB}{20,33,61}
\definecolor{charcoal}{RGB}{55,55,55}

\tikzset{
    databox/.style={
        rectangle, rounded corners=3pt,
        draw=charcoal!85, fill=white,
        thick,
        minimum width=2.8cm, minimum height=1.2cm,
        align=center, font=\sffamily,
        drop shadow
    },
    processbox/.style={
        rectangle, rounded corners=3pt,
        draw=charcoal!75, fill=gray!8,
        thick,
        minimum width=3.2cm, minimum height=1.2cm,
        align=center, font=\sffamily,
        drop shadow
    },
    networkbox/.style={
        rectangle, rounded corners=6pt,
        draw=gold!95!black,
        fill=navy!38,
        very thick,
        align=center, font=\sffamily,
        drop shadow
    },
    layer/.style={
        rectangle, rounded corners=2pt,
        minimum width=1cm, minimum height=2.5cm,
        font=\footnotesize\sffamily,
        drop shadow={opacity=0.25}
    },
    convlayer/.style={layer, draw=charcoal!85, fill=gray!18},
    relulayer/.style={layer, draw=gold!95!black, fill=gold!35},
    outputbox/.style={
        rectangle, rounded corners=3pt,
        draw=gold!85!black, fill=gold!15,
        thick,
        minimum width=2.8cm, minimum height=1.2cm,
        align=center, font=\sffamily,
        drop shadow
    },
    decision/.style={
        diamond,
        draw=navy!80, fill=navy!10,
        thick, aspect=2,
        minimum width=2.5cm, minimum height=1.5cm,
        align=center, font=\sffamily\small,
        drop shadow
    },
    circlebox/.style={
        circle,
        draw=charcoal!80, fill=gray!10,
        very thick,
        minimum size=3cm,
        align=center, font=\sffamily\small,
        drop shadow
    },
    arrow/.style={-Stealth, thick, draw=charcoal!80},
    thickarrow/.style={-Stealth, very thick, draw=gold!85!black},
    sectionlabel/.style={font=\sffamily\Large\bfseries, color=navy!90}
}

\def\BibTeX{{\rm B\kern-.05em{\sc i\kern-.025em b}\kern-.08em
    T\kern-.1667em\lower.7ex\hbox{E}\kern-.125emX}}

\begin{document}

\title{\emph{DopplerGLRTNet} for Radar Off-Grid Detection
\thanks{Part of this work was supported by ANR-ASTRID NEPTUNE 3 (ANR-23-ASM2-0009).}
}

\author{
\IEEEauthorblockN{
Y. A. Rouzoumka\IEEEauthorrefmark{1}\IEEEauthorrefmark{2}, 
J. Pinsolle\IEEEauthorrefmark{2},
E. Terreaux\IEEEauthorrefmark{1},
C. Morisseau\IEEEauthorrefmark{1},
J.-P. Ovarlez\IEEEauthorrefmark{1}\IEEEauthorrefmark{2} and
C. Ren\IEEEauthorrefmark{2}
}
\IEEEauthorblockA{
\IEEEauthorrefmark{1}\textit{DEMR, ONERA, Universit\'e Paris Saclay, F-91123 Palaiseau, France}, \\
\IEEEauthorrefmark{2}\textit{SONDRA, CentraleSupélec,  Université Paris-Saclay, 91190 Gif-sur-Yvette, France.}
}
}

\maketitle

\begin{abstract}
Off-grid targets whose Doppler (or angle) does not lie on the discrete processing grid can severely degrade classical normalized matched-filter (NMF) detectors: even at high SNR, the detection probability may saturate at operationally relevant low false-alarm rates.
A principled remedy is the continuous-parameter GLRT, which maximizes a normalized correlation over the parameter domain; however, dense scanning increases test-time cost and remains sensitive to covariance mismatch through whitening.
We propose \emph{DopplerGLRTNet}, an \emph{amortized off-grid GLRT}: a lightweight regressor predicts the continuous Doppler within a resolution cell from the whitened observation, and the detector outputs a \emph{single} GLRT/NMF-like score given by the normalized matched-filter energy at the predicted Doppler.
Monte Carlo simulations in Gaussian and compound-Gaussian clutter show that \emph{DopplerGLRTNet} mitigates off-grid saturation, approaches dense-scan performance at a fraction of its cost, and improves robustness to covariance estimation at the same empirically calibrated $P_{\mathrm{fa}}$.
\end{abstract}


\section{Introduction}
Detecting weak radar targets embedded in strong disturbance is a central problem in statistical signal processing.
In many scenarios, the target return is modeled as a (complex) steering vector corrupted by clutter and thermal noise, with unknown deterministic amplitude and nuisance parameters such as Doppler or angle.
Classical detection theory recommends generalized likelihood ratio tests (GLRTs), in which unknown parameters are replaced by their maximum-likelihood estimates and plugged into the likelihood ratio~\cite{scharf_statistical,kay1998_detection}.

\paragraph*{From GLRT to adaptive matched filtering}
When the steering vector is known up to an unknown complex amplitude and the disturbance is Gaussian with unknown power and known covariance shape, the GLRT reduces to the \emph{Normalized Matched Filter} (NMF).
In adaptive settings, closely related detectors (e.g., AMF/ANMF families) and their CFAR properties have been extensively studied~\cite{kelly1986adaptive,robey1992cfar_amf}.
Beyond Gaussian noise, NMF-type detectors remain central in complex elliptically symmetric and compound-Gaussian (SIRV) clutter models~\cite{conte1995compoundgaussian, de2016modern}, and robust scatter estimation (e.g., Tyler's fixed-point estimator) is a standard tool to improve robustness under heavy tails and mismatch~\cite{tyler1987scatter}.

\paragraph*{The off-grid problem: mismatch-induced saturation}
Practical implementations typically test Doppler/angle on a discrete grid (e.g., FFT bins).
Targets are rarely perfectly "on-grid", and even a small mismatch between the true and tested steering vectors can dramatically impair detection.
In particular, for very low operational false-alarm probabilities (e.g.\ $P_{\mathrm{fa}}\le 10^{-6}$), on-grid NMF can exhibit a saturation of $P_{\mathrm{d}}$ as SNR increases~\cite{develter2024offgridNMF}.

\paragraph*{Continuous-parameter GLRT and geometric characterization}
A principled remedy is to abandon the grid approximation and perform the GLRT over the \emph{continuous} parameter domain, yielding an \emph{off-grid} (continuous-parameter) NMF that maximizes a normalized correlation over a Doppler cell.
Formally, this is a test where the parameter is only present under $H_1$, and the resulting statistic is a supremum over a correlated continuum~\cite{davies1987nuisance}.
Recent work characterized the low-$P_{\mathrm{fa}}$ threshold relationship of this off-grid GLRT using a geometrical tube approximation on the unit sphere under compound-Gaussian noise and known covariance, clarifying validity conditions and the origin of saturation~\cite{develter2024offgridNMF}; see also the broader random-field geometry perspective in~\cite{adler2007randomfields}.
In practice, however, continuous maximization is implemented by dense scanning, which increases test-time complexity and remains sensitive to robustness to covariance estimation through whitening (and thus to the quality of the covariance/scatter estimate)~\cite{tyler1987scatter,conte1995compoundgaussian}.

\paragraph*{Learning-based detectors and structured neural approaches}
Learning-based detectors have been explored to complement CFAR-like pipelines.
For instance, \emph{CFARNet} enforces constant false-alarm behavior by design while leveraging deep architectures for target detection~\cite{diskin2024cfarnet}.
More broadly, ``model-driven'' deep networks and algorithm unrolling provide a systematic way to embed classical signal-processing structure into trainable architectures~\cite{monga2021unrolling,gregor2010lista}.
Learnable front-ends that preserve interpretability have also been popular in other wave settings (e.g., learnable filterbanks)~\cite{ravanelli2018sincnet}.
Finally, generative-model approaches have been investigated for radar detection under complex clutter, including VAE-based strategies for compound clutter and thermal noise~\cite{VAEalexis,rouzoumka2025complex}.

\paragraph*{Amortize the off-grid maximization}
Motivated by the off-grid mechanism, we propose a detector that preserves the GLRT/NMF approach while avoiding dense scanning:
instead of evaluating many Doppler hypotheses, we \emph{learn to predict} the continuous Doppler within a resolution cell and then evaluate a \emph{single} normalized matched filter at that predicted Doppler, which can be interpreted as an \emph{amortized} GLRT, in the sense that the maximization over Doppler is learned once and reused at inference time.
Importantly, the predictor can be purely real-valued , while the final detection statistic remains the standard complex normalized correlation energy.

\paragraph*{Contributions}
We introduce \emph{DopplerGLRTNet}, an amortized off-grid GLRT for Doppler off-grid detection:
(i) we formalize practical off-grid baselines (single-bin NMF and dense local scans) in the whitened-and-normalized domain~\cite{scharf_statistical,kay1998_detection,develter2024offgridNMF};
(ii) we propose a lightweight Doppler regressor constrained to a resolution cell, followed by a single GLRT-like normalized correlation score;
(iii) we demonstrate via Monte Carlo simulations in Gaussian and compound-Gaussian clutter that \emph{DopplerGLRTNet} mitigates off-grid saturation and approaches dense scanning at significantly reduced test-time cost.

\indent \textit{Notations}: Italic type indicates a scalar quantity, lower case boldface indicates a vector quantity, and upper case boldface indicates a matrix or tensor. The transpose and conjugate transpose operators are $^{\mathsf T}$ and $^{\mathsf H}$ respectively. $\mathbf{x}\sim\mathcal{CN}(\boldsymbol{\mu},\boldsymbol{\Sigma})$ is a  random complex circular Gaussian vector with mean vector $\boldsymbol{\mu}$ and covariance matrix $\boldsymbol{\Sigma}$. $\boldsymbol{\Sigma}^{1/2}$ denote any Hermitian square root of $\boldsymbol{\Sigma}$. $\mathcal{U}([a,b])$ is the uniform random distribution on $[a,b]$ interval.
$\lfloor .\rfloor$ denotes the floor function. $\mathbb{I}_{\{\cdot\}}$ denotes the indicator function.

\section{Problem Formulation}
\label{sec:problem}

\subsection{Signal Model in Slow-Time}
We consider a single range cell after range compression and focus on slow-time samples over a coherent processing interval of $m$ pulses. The received data form a complex vector
\begin{equation}
  \mathbf{z} = [z_0, \ldots, z_{m-1}]^{\mathsf T} \in \mathbb{C}^m,
\end{equation}
with $m=16$ in our main experiments.

We adopt the usual decomposition into clutter, thermal noise, and (possibly) a point-like target with complex steering vector $\mathbf{p}(\theta_0)$:
\begin{equation}
\label{eq:hypotheses}
\begin{cases}
  H_0: & \mathbf{z} = \mathbf{c} + \mathbf{n}, \\[0.2em]
  H_1: & \mathbf{z} = \alpha\,\mathbf{p}(\theta_0) + \mathbf{c} + \mathbf{n},
\end{cases}
\end{equation}
where $\alpha\in\mathbb{C}$ is an unknown deterministic amplitude and $\theta_0$ is a normalized Doppler frequency. We model the disturbance as either Gaussian clutter or compound-Gaussian clutter~\cite{conte1995compoundgaussian,tyler1987scatter}.
In the Gaussian case,
\begin{equation}
  \mathbf{c} \sim \mathcal{CN}(\mathbf{0},\boldsymbol{\Sigma}_{\mathrm{c}}), \qquad
  \mathbf{n} \sim \mathcal{CN}(\mathbf{0},\sigma^2 \mathbf{I}_m),
\end{equation}
so the total disturbance covariance is $\boldsymbol{\Sigma}=\boldsymbol{\Sigma}_{\mathrm{c}}+\sigma^2\mathbf{I}_m$.
In the SIRV case, $\mathbf{c}=\sqrt{\gamma}\,\mathbf{g}$ with $\mathbf{g}\sim\mathcal{CN}(\mathbf{0},\boldsymbol{\Sigma}_{\mathrm{c}})$ and $\gamma>0$ independent, i.e.,
$\mathbf{c}\mid\gamma \sim \mathcal{CN}(\mathbf{0},\gamma\,\boldsymbol{\Sigma}_{\mathrm{c}})$ and $\mathbb{E}[\gamma]=1$.

\subsection{Doppler Parametrization: On-Grid vs Off-Grid}
The Doppler steering vector is parameterized by a normalized frequency $\theta\in[0,1)$:
\begin{equation}
\label{eq:steering}
  \mathbf{p}(\theta)
  = \frac{1}{\sqrt{m}}
    \big[1, e^{j 2\pi \theta}, \ldots, e^{j 2\pi (m-1)\theta}\big]^{\mathsf T}
  \in \mathbb{C}^m.
\end{equation}
In a conventional Doppler filter bank, $\theta$ is tested on the FFT grid
\begin{equation}
  \mathcal{G}
  = \Big\{\theta_k = \tfrac{k}{m}\; :\; k=0,\ldots,m-1 \Big\},
\end{equation}
and the Doppler axis is partitioned into resolution cells
\begin{equation}
\label{eq:cells}
  D_k
  = \Big[\tfrac{k}{m} - \tfrac{1}{2m};\; \tfrac{k}{m} + \tfrac{1}{2m}\Big],
  \quad k=0,\ldots,m-1.
\end{equation}
Targets are generally off-grid: $\theta_0$ is continuous and may fall anywhere in a cell. We fix one cell $D=D_{k_0}$ (typically $k_0=0$) and draw $\theta_0\sim\mathcal{U}(D)$ to model off-grid targets.

\paragraph*{Off-grid mismatch and Doppler leakage}
Although the target is a single complex sinusoid in slow-time, when projected onto a discrete Doppler filter bank (FFT grid) an off-grid frequency produces a \emph{mismatch} with any single on-grid template. Equivalently, in the Doppler-FFT domain, the target energy \emph{leaks} across multiple bins.
In this work, we adopt a \emph{single-cell} setting: we fix one resolution cell $D=D_{k_0}$ and draw $\theta_0\sim\mathcal{U}(D)$, so the target is always off-grid \emph{within that cell}. The performance drop of on-grid NMF then comes from the mismatch between the true Doppler $\theta_0$ and the tested on-grid Doppler, which is fixed to the cell center $\theta_c=k_0/m$.

\subsection{Matched-Filter SNR and Whitening}
We define the SNR (in the whitened domain) as
\begin{equation}
\label{eq:snr_def}
  \mathrm{SNR} 
  = |\alpha|^2\,\mathbf{p}^{\mathsf H}(\theta_0)\, \boldsymbol{\Sigma}^{-1}\mathbf{p}(\theta_0).
\end{equation}
In simulations we set $\sigma^2=1$ and parameterize $\alpha \in \mathbb{C}$ to match a desired SNR (in dB):
\begin{equation}
  |\alpha|^2
  = 10^{\mathrm{SNR}/10}\,
    \left(\mathbf{p}^{\mathsf H}(\theta_0)\,\boldsymbol{\Sigma}^{-1}\,\mathbf{p}(\theta_0)\right)^{-1},
\end{equation}
with $\arg(\alpha)\sim\mathcal{U}([0,2\pi))$.

To connect with the geometrical analysis of~\cite{develter2024offgridNMF, adler2007randomfields}, we define whitened and energy-normalized observations:
\begin{equation}
\label{eq:whiten_norm}
  \mathbf{x} = \widehat{\boldsymbol{\Sigma}}^{-1/2}\mathbf{z}, \quad
  \mathbf{u} = \frac{\mathbf{x}}{\left\|\mathbf{x}\right\|_2},
  \quad
  \mathbf{v}(\theta)
  = \frac{\widehat{\boldsymbol{\Sigma}}^{-1/2}\mathbf{p}(\theta)}
         {\|\widehat{\boldsymbol{\Sigma}}^{-1/2}\mathbf{p}(\theta)\|_2},
\end{equation}
In our experiments, $\widehat{\boldsymbol{\Sigma}}$ is a global SCM estimate fitted on $H_0$ training data (or $\widehat{\boldsymbol{\Sigma}}=\mathbf{I}_m$ when no whitening is applied); only the oracle uses the true base covariance $\boldsymbol{\Sigma}_0$.

\begin{figure*}[!t]
  \centering
  {\tikzset{every picture/.style={scale=0.70, transform shape}}%
   \begin{tikzpicture}[node distance=1.5cm and 1cm]

\node[sectionlabel] (trainlabel) at (0, 8.5) {Training:};

\node[databox] (radar) at (0, 6.5) {Radar Data\\[2pt] $\mathbf{z} \in \mathbb{C}^m$};
\node[databox, below=0.8cm of radar] (h0h1) {$H_0$ / $H_1$\\[2pt] $\theta_0 \sim \mathcal{U}(D)$};

\node[processbox, right=2cm of radar, minimum width=3.5cm] (preprocess)
{Whitening \& Normalization\\[2pt]
$\mathbf{u} = \frac{\widehat{\boldsymbol{\Sigma}}^{-1/2}\mathbf{z}}{\|\widehat{\boldsymbol{\Sigma}}^{-1/2}\mathbf{z}\|_2}$};

\coordinate (netstart) at ($(preprocess.east) + (1,0)$);


\node[convlayer, anchor=west] (conv1) at (netstart) {\shortstack{1D Conv\\Layer}};
\node[relulayer, right=0.35cm of conv1] (relu1) {SiLU};
\node[convlayer, right=0.35cm of relu1] (conv2) {\shortstack{1D Conv\\Layer}};
\node[relulayer, right=0.35cm of conv2] (relu2) {SiLU};
\node[convlayer, right=0.35cm of relu2] (fc) {\shortstack{Fully\\Connected\\Layer}};

\draw[arrow] (conv1) -- (relu1);
\draw[arrow] (relu1) -- (conv2);
\draw[arrow] (conv2) -- (relu2);
\draw[arrow] (relu2) -- (fc);

\begin{scope}[on background layer]
  \node[networkbox,
        fit=(conv1)(relu1)(conv2)(relu2)(fc),
        inner sep=10pt,
        minimum width=0pt,
        minimum height=0pt] (network) {};
\end{scope}

\node[font=\sffamily\large\bfseries, color=gold!95, anchor=south]
  at (network.north) {Neural Doppler Regressor $g_{\boldsymbol{\phi}}$};

\node[below=0.22cm of network.south, font=\normalsize, align=center] (loss)
{$\mathcal{L}(\boldsymbol{\phi}) = \mathcal{L}_{\mathrm{CE}}(\boldsymbol{\phi}) + \lambda \mathcal{L}_{\delta}(\boldsymbol{\phi})$, where
$\mathcal{L}_{\delta} = \mathbb{E}\left[\mathrm{Huber}_\kappa\left(\widehat{\delta}(\mathbf{u}; \boldsymbol{\phi}), \delta\right) \; \mathbb{I}_{\{y=1\}}\ \right]$};

\node[outputbox, right=4.4cm of network.east, minimum width=3cm] (theta_pred)
{Predicted Doppler\\[2pt] $\widehat{\theta}(\mathbf{u}; \boldsymbol{\phi}) \in D$};

\draw[arrow] (radar) -- (preprocess);
\draw[arrow] (h0h1) -| (preprocess);
\draw[arrow] (preprocess.east) -- (conv1.west);

\draw[arrow] (fc.east) --
  node[midway, above=2pt, align=center, text width=4cm]
  {$\widehat{\delta}(\mathbf{u}; \boldsymbol{\phi})) = \tanh(g(\mathbf{u}; \boldsymbol{\phi}))$}
  (theta_pred.west);

\node[sectionlabel, anchor=west] (detlabel) at ($(radar.west)+(0,-4.0cm)$) {Detection:};
\node[databox, anchor=west] (testdata)  at ($(radar.west)+(0,-6.0cm)$) {Test Radar\\Data\\[2pt] $\mathbf{z}$};

\node[circlebox, right=2cm of testdata, minimum size=3.5cm] (glrt)
{amortized GLRT\\[4pt] $T(\mathbf{u}; \boldsymbol{\phi}) =$\\[2pt]
$\left|\mathbf{v}(\widehat{\theta}(\mathbf{u}; \boldsymbol{\phi}))^{\mathsf{H}} \mathbf{u}\right|^2$};

\node[processbox, right=2cm of glrt, minimum width=3.8cm] (calibscore)
{Calibration (CFAR)\\[2pt]
$\tau_{P_{\mathrm{fa}}}=\mathrm{Quantile}_{1-P_{\mathrm{fa}}}\!\Big(\{T(\mathbf{u};\boldsymbol{\phi})\}_{H_0}\Big)$};


\node[decision, right=2cm of calibscore] (decision)
{$T(\mathbf{u}; \boldsymbol{\phi})> \tau_{P_{\mathrm{fa}}}$};

\node[outputbox, below=1.2cm of decision] (h1) {Target\\Present\\[2pt] $H_1$};
\node[outputbox, right=1.8cm of decision, minimum width=2.6cm] (h0)
{No Target\\[2pt] $H_0$};

\draw[arrow] (testdata) -- node[above, font=\scriptsize] {Preprocess} (glrt);
\draw[thickarrow] (glrt) -- (calibscore);
\draw[arrow] (calibscore) -- (decision);
\draw[arrow] (decision) -- node[right, font=\scriptsize] {Yes} (h1);
\draw[arrow] (decision.east) -- node[above, font=\scriptsize] {No} (h0.west);

\draw[thickarrow, dashed, gold!85!black, bend left=15]
  (theta_pred.south)
  to node[right, font=\scriptsize, align=left, yshift=-12pt] {Frozen\\network}
  ($(glrt.north) + (0, 0.05)$);

\begin{scope}[on background layer]
    \node[draw=gray!55, fill=gray!3, rounded corners=8pt, thick, dashed, 
          fit=(trainlabel) (radar) (h0h1) (preprocess) (network) (theta_pred) (loss),
          inner sep=10pt] {};
    \node[draw=gray!55, fill=gray!3, rounded corners=8pt, thick, dashed,
          fit=(detlabel) (testdata) (glrt) (calibscore) (decision) (h1) (h0),
          inner sep=10pt] {};
\end{scope}

\end{tikzpicture}%
  }
  \caption{Overview of \emph{DopplerGLRTNet}. \emph{Training} (top): a cell-constrained Doppler regressor outputs $\widehat{\theta}(\mathbf{u})\in D$ from whitened and energy-normalized slow-time data, optimized with $\mathcal{L}=\mathcal{L}_{\mathrm{CE}}+\lambda \mathcal{L}_{\delta}$ (Huber loss on the normalized offset for $H_1$ samples). \emph{Detection} (bottom): the frozen network predicts $\widehat{\theta}$ and we compute a single GLRT/NMF-like score $T(\mathbf{u}; \boldsymbol{\phi})=\big|\mathbf{v}^{\mathsf H}(\widehat{\theta}(\mathbf{u}; \boldsymbol{\phi}))\, \mathbf{u}\big|^2$, followed by an empirical CFAR threshold calibration $\tau_{P_{fa}}$.} 
  \label{fig:thetaglrtnet_pipeline}
\end{figure*}
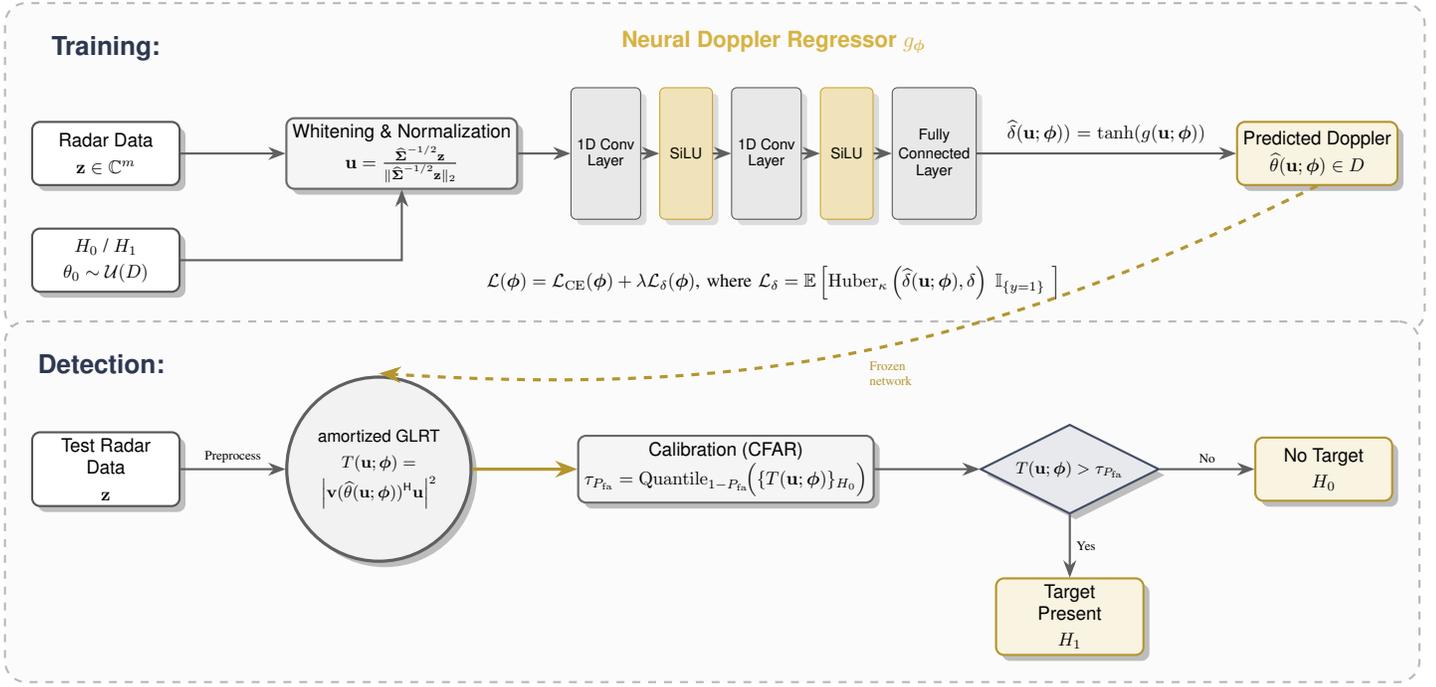

\section{Classical GLRT-Based Detectors}
\label{sec:classical}

\subsection{Normalized Matched Filter (Single Test)}
For a fixed steering vector $\mathbf{p}(\theta)$, the NMF statistic is~\cite{scharf_statistical,kay1998_detection,develter2024offgridNMF}
\begin{equation}
\label{eq:nmf}
  T_{\mathrm{NMF}}(\mathbf{z};\theta)
  = \frac{\left|\mathbf{p}^{\mathsf H}(\theta)\, \boldsymbol{\Sigma}^{-1}\, \mathbf{z}\right|^2}
         {\left(\mathbf{p}^{\mathsf H}(\theta)\, \boldsymbol{\Sigma}^{-1}\, \mathbf{p}(\theta)\right)
          \left(\mathbf{z}^{\mathsf H}\,\boldsymbol{\Sigma}^{-1}\,\mathbf{z}\right)}.
\end{equation}
In whitened and normalized coordinates \eqref{eq:whiten_norm}, this reduces to
\begin{equation}
\label{eq:nmf_u}
  T_{\mathrm{NMF}}(\mathbf{u};\theta)
  = \big|\mathbf{v}^{\mathsf H}(\theta)\,\mathbf{u}\big|^2 \in [0;1].
\end{equation}
Under $H_0$ and exact whitening, $T_{\mathrm{NMF}}(\mathbf{u};\theta)$ follows a Beta distribution with parameters $(1,m-1)$, yielding
\begin{equation}
\label{eq:pfa_nmf_single}
  P_{\mathrm{fa}}
  = \mathbb{P}\big(T_{\mathrm{NMF}}(\mathbf{u};\theta) > w^2 \mid H_0\big)
  = (1 - w^2)^{m-1}.
\end{equation}
This closed-form relationship holds for a \emph{single} tested $\theta$.

\subsection{On-grid NMF (single bin) and Local Dense Scanning (NMF scan within one cell)}
In the single-cell setting (fixed $D=D_{k_0}$), an on-grid baseline tests only the cell center $\theta_{k_0}=k_0/m$:
\[
T_{\mathrm{NMF}}(\mathbf{u};\theta_{k_0}) = \big|\mathbf{v}^{\mathsf H}(\theta_{k_0})\, \mathbf{u}\big|^2.
\]
A stronger baseline numerically approximates the off-grid GLRT by \emph{densely scanning only inside the cell}:
\begin{equation}
\label{eq:local_scan_grid}
\mathcal{G}_{L}(D_{k_0})
= \big\{\theta^{(\ell)}\big\}_{\ell=1}^{K}
\subset D_{k_0},
\,\,
\theta^{(\ell)} \!= \theta_{k_0} + \tfrac{\xi_\ell}{2m}\,,
\,\,
\xi_\ell \in [-1;1],
\end{equation}
with \ $K=64$ points uniformly spaced in $D_{k_0}$. The resulting \emph{local scan} statistic is
\begin{equation}
\label{eq:nmf_scan_local}
T_{\mathrm{NMF}}^{\max}(\mathbf{u};D_{k_0})
= \max_{\theta\in \mathcal{G}_{L}(D_{k_0})}
\big|\mathbf{v}^{\mathsf H}(\theta)\mathbf{u}\big|^2,
\end{equation}
which provides a practical numerical approximation to the continuous off-grid GLRT over $D_{k_0}$, at the cost of $K$ correlations per tested vector.
Since these tests are correlated, the global $P_{\mathrm{fa}}$ of the maximum is not given by the single-test formula. Therefore, we calibrate the thresholds empirically under $H_0$ for each detector (see Section~\ref{sec:training}).

\subsection{Off-Grid GLRT Viewpoint (Continuous-Parameter NMF)}
A principled way to mitigate off-grid mismatch is to perform the GLRT over the continuous Doppler interval $D$:
\begin{equation}
\label{eq:offgrid_glrt}
  T_{\mathrm{GLRT}}(\mathbf{u}; D)
  = \max_{\theta \in D} \big|\mathbf{v}^{\mathsf H}(\theta)\, \mathbf{u}\big|^2.
\end{equation}
Its null distribution under $H_0$ is nontrivial (it corresponds to the maximum over a correlated continuum), and tube formula approximations can be derived in the low-$P_{\mathrm{fa}}$ regime~\cite{develter2024offgridNMF, Trottier2025OffGridRD}. In practice, Equation \eqref{eq:nmf_scan_local} provides a numerical approximation to Equation \eqref{eq:offgrid_glrt} when $\mathcal{G}_{L}(D)\subset D$ is sufficiently dense.

\subsection{Oracle Upper Bound}
To quantify an upper bound, we use a genie-aided oracle that has access to both the true Doppler $\theta_0$ and the true covariance $\boldsymbol{\Sigma}$ (matched whitening). In the whitened-and-normalized domain, the oracle score is
\begin{align}
\label{eq:oracle_theta_true_trueSigma}
T_{\mathrm{oracle}}(\mathbf{u};\theta_0)
&=
\left|
\left(
\frac{\boldsymbol{\Sigma}^{-1/2}\mathbf{p}(\theta_0)}
     {\big\|\boldsymbol{\Sigma}^{-1/2}\mathbf{p}(\theta_0)\big\|_2}
\right)^{\mathsf H}\!\!\!
\left(
\frac{\boldsymbol{\Sigma}^{-1/2}\mathbf{z}}
     {\big\|\boldsymbol{\Sigma}^{-1/2}\mathbf{z}\big\|_2}
\right)
\right|^2 \, ,\nonumber\\
&=
\big|\mathbf{v}_{\mathrm{true}}^{\mathsf H}(\theta_0)\, \mathbf{u}_{\mathrm{true}}\big|^2 .
\end{align}
This oracle is not implementable in practice but provides a meaningful performance ceiling for off-grid detection.

\section{\emph{DopplerGLRTNet}: Neural Off-Grid detector via Amortized GLRT}
\label{sec:neural_glrt}

\emph{DopplerGLRTNet} targets the continuous maximization in the off-grid GLRT~\eqref{eq:offgrid_glrt}.
Instead of dense scanning~\eqref{eq:nmf_scan_local} (many templates per test vector), it learns an \emph{amortized} GLRT:
\[
\mathbf{u}\mapsto \widehat{\theta}(\mathbf{u})\in D
\quad\Rightarrow\quad
\left|\mathbf{v}^{\mathsf H}\left(\widehat{\theta}(\mathbf{u})\right)\mathbf{u}\right|^2.
\]

\paragraph*{Cell-constrained Doppler prediction}
Let $D=[\theta_c-\frac{1}{2m};\theta_c+\frac{1}{2m}]$. The network predicts a normalized offset $\widehat{\delta}(\mathbf{u})\in[-1;1]$ and maps it to a continuous Doppler inside the cell:
\begin{equation}
\label{eq:theta_hat_param}
\widehat{\theta}(\mathbf{u}; \boldsymbol{\phi})=\theta_c+\tfrac{1}{2m}\,\widehat{\delta}(\mathbf{u};\boldsymbol{\phi}),
\;
\widehat{\delta}(\mathbf{u}; \boldsymbol{\phi})=\tanh\!\big(g(\mathbf{u}; {\boldsymbol{\phi}})\big),
\end{equation}
where $g(.; {\boldsymbol{\phi}})$ operates on two real channels
$\mathbf{x}_{\mathbb{R}}=[\Re\{\mathbf{x}\},\Im\{\mathbf{x}\}]\in\mathbb{R}^{2m}$, while detection uses complex-valued templates and complex inner products. ${\boldsymbol{\phi}}$ are the learnable parameters of the neural Doppler regressor.

\paragraph*{One-correlation GLRT-like score}
Given $\widehat{\theta}(\mathbf{u};\boldsymbol{\phi})$ and using the normalizations of \eqref{eq:whiten_norm},
we evaluate the normalized correlation energy:

\begin{equation}
\label{eq:theta_score_core}
T(\mathbf{u};\boldsymbol{\phi}) =
 \left|\mathbf{v}^{\mathsf H}\left(\widehat{\theta}(\mathbf{u};\boldsymbol{\phi})\right)\mathbf{u}\right|^2\in[0;1]
\end{equation}
A threshold on $T(\mathbf{u};\boldsymbol{\phi})$ is set by empirical $H_0$ calibration (see Section~\ref{sec:training}).

\paragraph*{Complexity}
\emph{DopplerGLRTNet} uses one length-$m$ complex inner product per test vector (vs.\ $K$ correlations for scanning) plus one template generation $\widehat{\boldsymbol{\Sigma}}^{-1/2}\mathbf{p}(\widehat{\theta})$, negligible for $m\in\{16,32\}$ and still far below dense scanning.

\section{Training, Calibration, and Simulation Setup}
\label{sec:training}

We follow standard radar evaluation protocol: generate $H_0/H_1$ samples, calibrate thresholds on independent $H_0$ data to match a desired $P_{\mathrm{fa}}$, and then estimate $P_{\mathrm{d}}$ versus SNR on $H_1$ test data.

\paragraph*{Disturbance model and whitening}
Clutter covariance follows a Toeplitz AR(1) model, with  $[\boldsymbol{\Sigma}_{\mathrm{c}}]_{i,j}=\rho^{|i-j|}$, and additive thermal Gaussian noise $\sigma^2\, \mathbf{I}_m$.
Compound-Gaussian clutter is generated as  $\mathbf{c}=\sqrt{\gamma}\,\mathbf{g}$, where $\mathbf{g}\sim\mathcal{CN}(\mathbf{0},\boldsymbol{\Sigma}_{\mathrm{c}})$. The texture component $\gamma$ is
sampled from a Gamma distribution $\Gamma(\mu, \frac{1}{\mu})$ with $\mu = 1$.
All detectors operate on the same whitened and energy-normalized vector (cf.\ \eqref{eq:whiten_norm}); in adaptive simulations, $\widehat{\boldsymbol{\Sigma}}$ is estimated from $H_0$ training data (e.g. SCM or Tyler).

\paragraph*{Single-cell off-grid injection}
We fix a cell $D=D_{k_0}$ and draw $\theta_0\sim\mathcal{U}(D_{k_0})$ (continuous, no discretization). The data are generated as
\[
\mathbf{z}_{H_0}=\mathbf{c}+\mathbf{n},
\qquad
\mathbf{z}_{H_1}=\alpha\,\mathbf{p}(\theta_0)+\mathbf{c}+\mathbf{n},
\]
where $\arg(\alpha)\sim\mathcal{U}([0;2\pi))$ and $|\alpha|$ is set to match the MF-SNR using the base covariance
$\boldsymbol{\Sigma}=\boldsymbol{\Sigma}_{\mathrm{c}}+\sigma^2\,\mathbf{I}_m$:
\begin{equation}
|\alpha|^2
=10^{\mathrm{SNR}/10}\,
\left(\mathbf{p}^{\mathsf H}(\theta_0)\,\boldsymbol{\Sigma}^{-1}\,\mathbf{p}(\theta_0)\right)^{-1}.
\end{equation}
When evaluating multiple cells, we train one \emph{DopplerGLRTNet} per $D_k$ with $\theta_c=k/m$ using the same $\theta_0\sim\mathcal{U}(D_k)$ protocol.

\paragraph*{Splits, objective, and calibration}
We use (i) a supervised train/validation mixture of $H_0/H_1$ samples across SNR levels to learn $\boldsymbol{\phi}$, and (ii) an independent $H_0$ calibration set for threshold selection.

We optimize the standard logistic cross-entropy:
\begin{equation*}
\label{eq:ce_loss}
\mathcal{L}_{\mathrm{CE}}(\boldsymbol{\phi})
= - \mathbb{E} \Big[
 y \log\big(T(\mathbf{u};\boldsymbol{\phi})\big)
 + (1-y)\log\big(1-T(\mathbf{u};\boldsymbol{\phi})\big)
\Big],
\end{equation*}

\emph{Huber regression for Doppler:}
In the $H_1$ case ($y=1$), we supervise the normalized Doppler offset
$\delta=2m(\theta_0-\theta_c)\in[-1;1]$ using the Huber loss~\cite{huber2009robust},
which is quadratic for small errors and linear for large errors:
\begin{equation}
\label{eq:huber_def}
\mathrm{Huber}_\kappa(e)=
\begin{cases}
\frac{1}{2}e^2, & |e|\le \kappa,\\
\kappa(|e|-\frac{1}{2}\kappa), & |e|>\kappa\, .
\end{cases}
\end{equation}
This yields
\begin{equation}
\label{eq:theta_reg_loss}
\mathcal{L}_{\delta}(\boldsymbol{\phi})
=
\mathbb{E}\Big[
\mathrm{Huber}_\kappa\big(\widehat{\delta}(\mathbf{u}; \boldsymbol{\phi})-\delta\big) \; \mathbb{I}_{\{y=1\}}
\Big].
\end{equation}

The final objective is
\begin{equation}
\label{eq:total_loss}
\begin{aligned}
\mathcal{L}(\boldsymbol{\phi})
&=
\underbrace{\mathcal{L}_{\mathrm{CE}}(\boldsymbol{\phi})}_{\text{logistic detection loss}}
+ \;
\lambda\,
\underbrace{\mathcal{L}_{\delta}(\boldsymbol{\phi})}_{\text{Huber Doppler loss}} .
\end{aligned}
\end{equation}

After training, we freeze parameters and we set $\tau_{P_{\mathrm{fa}}}$ as the empirical $(1-P_{\mathrm{fa}})$-quantile of $\{T(\mathbf{u};\boldsymbol{\phi})\}_{H_0}$; we decide $H_1$ if $T(\mathbf{u};\boldsymbol{\phi})>\tau_{P_{\mathrm{fa}}}$.
Crucially, \emph{all} detectors (classical and neural) are calibrated this way to enforce the same target $P_{\mathrm{fa}}$.

The overall structure of the approach is described in figure \ref{fig:thetaglrtnet_pipeline}. The neural doppler regressor is optimized using the objective loss \eqref{eq:total_loss} over 40 epochs with the Adam optimizer and a learning rate of $2 \times  10^{-3}$. The supervised training set contains $N_{\mathrm{train}}=10{,}000$ samples, split equally between $H_0$ and $H_1$,
and the validation set contains $N_{\mathrm{val}}=5{,}000$ samples, also balanced.
During supervised training, positive ($H_1$) samples are generated by sampling SNR values from the evaluation grid.


\section{Experimental Results}
\label{sec:experiments}

We evaluate \emph{DopplerGLRTNet} on simulated data under Gaussian and compound-Gaussian clutter, reporting $P_{\mathrm{d}}$ versus SNR at fixed calibrated $P_{\mathrm{fa}}$.

\pgfplotsset{
    tiny,
    legend style={
        at={(0.03,0.97)},
        anchor=north west,
        font=\scriptsize,
        draw=none,
        fill=white,
        fill opacity=0.85,
        text opacity=1,
    },
    every axis plot/.append style={
        line width=1.1pt,
    },
}

\begin{figure*}[!htb]
\centering
\begin{tikzpicture}
\begin{groupplot}[
    group style={
        group size=3 by 1,
        horizontal sep=0.6cm,
    },
    width=0.38\textwidth,
    height=55mm,
    xlabel={\small{SNR (dB)}},
    xlabel style={yshift=2pt},
    ylabel={\normalsize{$P_{\mathrm{d}}$}},
    ylabel style={yshift=-2pt},
    xmin=-20, xmax=20,
    ymin=-0.03, ymax=1.03,
    xtick={-20, -15,-10, -5, 0, 5, 10, 15, 20},
    ytick={0,0.2,0.4,0.6,0.8,1.0},
    tick label style={font=\scriptsize},
    legend columns=1,
    ymajorgrids=true,
    xmajorgrids=true,
    grid style={dashed, gray!25},
    axis line style={semithick},
]

\nextgroupplot[title={\small (a) cGN + AWGN}]

\addplot[color=blue, mark=none, loosely dashed] coordinates {
    (-20.0,0.0112)(-19.0,0.0116)(-18.0,0.0118)(-17.0,0.0118)(-16.0,0.013)
    (-15.0,0.012)(-14.0,0.0134)(-13.0,0.0132)(-12.0,0.0152)(-11.0,0.015)
    (-10.0,0.0148)(-9.0,0.0154)(-8.0,0.017)(-7.0,0.02)(-6.0,0.0206)
    (-5.0,0.0218)(-4.0,0.0262)(-3.0,0.0342)(-2.0,0.0366)(-1.0,0.0442)
    (0.0,0.0612)(1.0,0.0718)(2.0,0.0906)(3.0,0.1152)(4.0,0.1562)
    (5.0,0.2054)(6.0,0.269)(7.0,0.3576)(8.0,0.4598)(9.0,0.561)
    (10.0,0.6618)(11.0,0.7452)(12.0,0.8118)(13.0,0.8584)(14.0,0.899)
    (15.0,0.931)(16.0,0.9436)(17.0,0.9664)(18.0,0.9758)(19.0,0.9848)(20.0,0.9904)
};
\addlegendentry{NMF on-grid}

\addplot[color=red, mark=none, dotted] coordinates {
    (-20.0,0.0092)(-19.0,0.0088)(-18.0,0.0098)(-17.0,0.0088)(-16.0,0.0086)
    (-15.0,0.0092)(-14.0,0.0086)(-13.0,0.009)(-12.0,0.0092)(-11.0,0.0094)
    (-10.0,0.0098)(-9.0,0.011)(-8.0,0.011)(-7.0,0.0126)(-6.0,0.0148)
    (-5.0,0.0166)(-4.0,0.0194)(-3.0,0.0238)(-2.0,0.028)(-1.0,0.0348)
    (0.0,0.0454)(1.0,0.062)(2.0,0.0838)(3.0,0.1126)(4.0,0.1564)
    (5.0,0.2192)(6.0,0.3038)(7.0,0.4158)(8.0,0.5462)(9.0,0.684)
    (10.0,0.8066)(11.0,0.91)(12.0,0.9684)(13.0,0.9956)(14.0,0.9992)
    (15.0,0.9998)(16.0,1.0)(17.0,1.0)(18.0,1.0)(19.0,1.0)(20.0,1.0)
};
\addlegendentry{NMF scan}

\addplot[color=green!60!black, mark=none, dashed] coordinates {
    (-20.0,0.0094)(-19.0,0.0088)(-18.0,0.01)(-17.0,0.0088)(-16.0,0.009)
    (-15.0,0.0092)(-14.0,0.0088)(-13.0,0.0088)(-12.0,0.0094)(-11.0,0.0094)
    (-10.0,0.0098)(-9.0,0.0112)(-8.0,0.011)(-7.0,0.0126)(-6.0,0.0152)
    (-5.0,0.0168)(-4.0,0.0198)(-3.0,0.0238)(-2.0,0.0284)(-1.0,0.0352)
    (0.0,0.0466)(1.0,0.0628)(2.0,0.0834)(3.0,0.113)(4.0,0.1568)
    (5.0,0.2196)(6.0,0.3044)(7.0,0.4164)(8.0,0.5468)(9.0,0.684)
    (10.0,0.807)(11.0,0.9096)(12.0,0.9688)(13.0,0.9952)(14.0,0.9992)
    (15.0,0.9998)(16.0,1.0)(17.0,1.0)(18.0,1.0)(19.0,1.0)(20.0,1.0)
};
\addlegendentry{DopplerGLRTNet}

\addplot[color=violet, mark=none] coordinates {
    (-20.0,0.0096)(-19.0,0.0102)(-18.0,0.0076)(-17.0,0.0082)(-16.0,0.0104)
    (-15.0,0.0102)(-14.0,0.0096)(-13.0,0.0102)(-12.0,0.0102)(-11.0,0.0114)
    (-10.0,0.0118)(-9.0,0.0122)(-8.0,0.0134)(-7.0,0.0158)(-6.0,0.0176)
    (-5.0,0.0202)(-4.0,0.024)(-3.0,0.029)(-2.0,0.036)(-1.0,0.0452)
    (0.0,0.0584)(1.0,0.078)(2.0,0.1044)(3.0,0.1406)(4.0,0.1962)
    (5.0,0.2756)(6.0,0.3784)(7.0,0.5118)(8.0,0.6552)(9.0,0.7924)
    (10.0,0.8948)(11.0,0.9516)(12.0,0.9878)(13.0,0.9986)(14.0,1.0)
    (15.0,1.0)(16.0,1.0)(17.0,1.0)(18.0,1.0)(19.0,1.0)(20.0,1.0)
};
\addlegendentry{Oracle}

\nextgroupplot[title={\small (b) cCGN}, ylabel={}]

\addplot[color=blue, mark=none, loosely dashed] coordinates {
    (-20.0,0.02)(-19.0,0.024)(-18.0,0.0256)(-17.0,0.0288)(-16.0,0.0328)
    (-15.0,0.0364)(-14.0,0.0406)(-13.0,0.047)(-12.0,0.0498)(-11.0,0.0538)
    (-10.0,0.0648)(-9.0,0.0752)(-8.0,0.0834)(-7.0,0.0954)(-6.0,0.111)
    (-5.0,0.1312)(-4.0,0.1454)(-3.0,0.1646)(-2.0,0.1912)(-1.0,0.2152)
    (0.0,0.244)(1.0,0.2782)(2.0,0.3128)(3.0,0.3574)(4.0,0.3928)
    (5.0,0.4406)(6.0,0.4936)(7.0,0.5438)(8.0,0.5942)(9.0,0.6498)
    (10.0,0.701)(11.0,0.7444)(12.0,0.7846)(13.0,0.8394)(14.0,0.8712)
    (15.0,0.8958)(16.0,0.9162)(17.0,0.9416)(18.0,0.9578)(19.0,0.9616)(20.0,0.972)
};
\addlegendentry{NMF on-grid}

\addplot[color=red, mark=none, dotted] coordinates {
    (-20.0,0.0208)(-19.0,0.025)(-18.0,0.0262)(-17.0,0.0278)(-16.0,0.0316)
    (-15.0,0.0366)(-14.0,0.0402)(-13.0,0.0468)(-12.0,0.0498)(-11.0,0.0568)
    (-10.0,0.0632)(-9.0,0.0754)(-8.0,0.0862)(-7.0,0.099)(-6.0,0.1144)
    (-5.0,0.1348)(-4.0,0.1514)(-3.0,0.1722)(-2.0,0.1972)(-1.0,0.2244)
    (0.0,0.2532)(1.0,0.2864)(2.0,0.326)(3.0,0.368)(4.0,0.4116)
    (5.0,0.461)(6.0,0.515)(7.0,0.5698)(8.0,0.6264)(9.0,0.6816)
    (10.0,0.737)(11.0,0.7922)(12.0,0.8484)(13.0,0.904)(14.0,0.9472)
    (15.0,0.971)(16.0,0.982)(17.0,0.992)(18.0,0.9958)(19.0,0.9992)(20.0,0.9998)
};
\addlegendentry{NMF scan}

\addplot[color=green!60!black, mark=none, dashed] coordinates {
    (-20.0,0.0214)(-19.0,0.023)(-18.0,0.0254)(-17.0,0.0272)(-16.0,0.0302)
    (-15.0,0.035)(-14.0,0.0392)(-13.0,0.0466)(-12.0,0.0502)(-11.0,0.057)
    (-10.0,0.0642)(-9.0,0.076)(-8.0,0.087)(-7.0,0.1)(-6.0,0.1162)
    (-5.0,0.1368)(-4.0,0.153)(-3.0,0.174)(-2.0,0.2002)(-1.0,0.2276)
    (0.0,0.2562)(1.0,0.2896)(2.0,0.328)(3.0,0.3708)(4.0,0.4144)
    (5.0,0.4632)(6.0,0.5172)(7.0,0.573)(8.0,0.629)(9.0,0.685)
    (10.0,0.7396)(11.0,0.7946)(12.0,0.8506)(13.0,0.9046)(14.0,0.9478)
    (15.0,0.9718)(16.0,0.9812)(17.0,0.991)(18.0,0.9954)(19.0,0.9988)(20.0,0.9998)
};
\addlegendentry{DopplerGLRTNet}

\addplot[color=violet, mark=none] coordinates {
    (-20.0,0.0226)(-19.0,0.0246)(-18.0,0.0282)(-17.0,0.0306)(-16.0,0.0362)
    (-15.0,0.0392)(-14.0,0.0442)(-13.0,0.0512)(-12.0,0.055)(-11.0,0.0632)
    (-10.0,0.0716)(-9.0,0.084)(-8.0,0.0986)(-7.0,0.1138)(-6.0,0.1324)
    (-5.0,0.1562)(-4.0,0.173)(-3.0,0.196)(-2.0,0.2236)(-1.0,0.25)
    (0.0,0.2796)(1.0,0.316)(2.0,0.3568)(3.0,0.3992)(4.0,0.4444)
    (5.0,0.494)(6.0,0.548)(7.0,0.6022)(8.0,0.6616)(9.0,0.7176)
    (10.0,0.7738)(11.0,0.8236)(12.0,0.8712)(13.0,0.919)(14.0,0.9536)
    (15.0,0.9736)(16.0,0.9886)(17.0,0.9956)(18.0,0.9978)(19.0,0.9996)(20.0,1.0)
};
\addlegendentry{Oracle}

\nextgroupplot[title={\small (c) cCGN + AWGN}, ylabel={}]

\addplot[color=blue, mark=none, loosely dashed] coordinates {
    (-20.0,0.0072)(-19.0,0.0066)(-18.0,0.0066)(-17.0,0.0066)(-16.0,0.0066)
    (-15.0,0.0066)(-14.0,0.0078)(-13.0,0.0068)(-12.0,0.0058)(-11.0,0.0072)
    (-10.0,0.0092)(-9.0,0.0084)(-8.0,0.008)(-7.0,0.0128)(-6.0,0.0118)
    (-5.0,0.012)(-4.0,0.0156)(-3.0,0.0204)(-2.0,0.0286)(-1.0,0.0322)
    (0.0,0.043)(1.0,0.0606)(2.0,0.0796)(3.0,0.1126)(4.0,0.1556)
    (5.0,0.2292)(6.0,0.3052)(7.0,0.397)(8.0,0.4952)(9.0,0.5808)
    (10.0,0.6866)(11.0,0.756)(12.0,0.8082)(13.0,0.8492)(14.0,0.8966)
    (15.0,0.9224)(16.0,0.9366)(17.0,0.9618)(18.0,0.9722)(19.0,0.9804)(20.0,0.9878)
};
\addlegendentry{NMF on-grid}

\addplot[color=red, mark=none, dotted] coordinates {
    (-20.0,0.0074)(-19.0,0.0076)(-18.0,0.0072)(-17.0,0.0078)(-16.0,0.0072)
    (-15.0,0.0074)(-14.0,0.007)(-13.0,0.007)(-12.0,0.0074)(-11.0,0.0078)
    (-10.0,0.0086)(-9.0,0.0094)(-8.0,0.0104)(-7.0,0.0118)(-6.0,0.0138)
    (-5.0,0.0164)(-4.0,0.0198)(-3.0,0.0248)(-2.0,0.031)(-1.0,0.039)
    (0.0,0.0524)(1.0,0.0718)(2.0,0.0964)(3.0,0.1302)(4.0,0.179)
    (5.0,0.2482)(6.0,0.3372)(7.0,0.4486)(8.0,0.5792)(9.0,0.7116)
    (10.0,0.8244)(11.0,0.9076)(12.0,0.9598)(13.0,0.9862)(14.0,0.9952)
    (15.0,0.999)(16.0,0.9992)(17.0,1.0)(18.0,1.0)(19.0,1.0)(20.0,1.0)
};
\addlegendentry{NMF scan}

\addplot[color=green!60!black, mark=none, dashed] coordinates {
    (-20.0,0.0066)(-19.0,0.0066)(-18.0,0.0068)(-17.0,0.0068)(-16.0,0.0064)
    (-15.0,0.0066)(-14.0,0.0066)(-13.0,0.0066)(-12.0,0.007)(-11.0,0.007)
    (-10.0,0.0078)(-9.0,0.0086)(-8.0,0.0094)(-7.0,0.0112)(-6.0,0.0134)
    (-5.0,0.0158)(-4.0,0.0196)(-3.0,0.0246)(-2.0,0.031)(-1.0,0.0392)
    (0.0,0.0526)(1.0,0.0724)(2.0,0.097)(3.0,0.1316)(4.0,0.1798)
    (5.0,0.2486)(6.0,0.3368)(7.0,0.4484)(8.0,0.5794)(9.0,0.7118)
    (10.0,0.8246)(11.0,0.907)(12.0,0.96)(13.0,0.9864)(14.0,0.995)
    (15.0,0.9992)(16.0,0.999)(17.0,1.0)(18.0,1.0)(19.0,1.0)(20.0,1.0)
};
\addlegendentry{DopplerGLRTNet}

\end{groupplot}
\end{tikzpicture}
\caption{$P_{\mathrm{d}}$ versus matched-filter SNR at $P_{\mathrm{fa}}=10^{-2}$ for off-grid targets in the single Doppler cell $D_0$ ($m=16$, $\rho=0.5$). Whitening uses a global SCM fit on $H_0$ data, followed by energy normalization. Baselines: on-grid NMF, local scan in $D_0$ ($K=64$), oracle, and \emph{DopplerGLRTNet}. (cGN: complex Gaussian clutter; cCGN: compound-Gaussian clutter; AWGN: additive white Gaussian noise.)}
\label{fig:detection_scenarios}
\end{figure*}
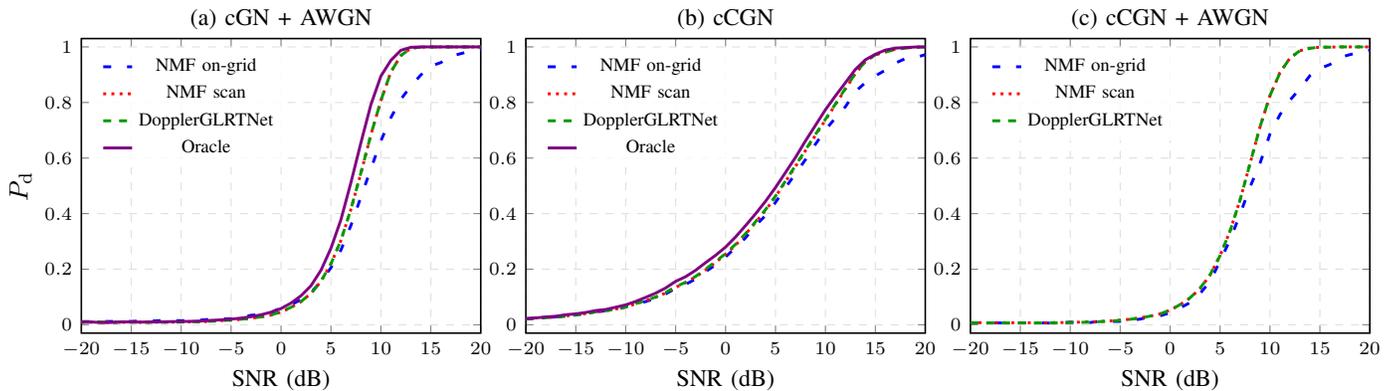


Fig.~\ref{fig:detection_scenarios} reports $P_{\mathrm{d}}$ versus SNR at calibrated $P_{\mathrm{fa}}=10^{-2}$ for three disturbance settings:
(a) complex Gaussian clutter with additive white Gaussian noise, using global SCM whitening estimated on $H_0$ training data,
(b) compound-Gaussian clutter,
and (c) compound-Gaussian clutter with additional AWGN. 
Across all methods, thresholds are calibrated on independent $H_0$ data using the same empirical quantile rule.

\subsection{Baseline Detectors}
\label{subsec:baselines}
We compare on-grid NMF at the cell center $\theta_c$~\cite{scharf_statistical,kay1998_detection,develter2024offgridNMF},
dense local NMF scan over $\mathcal{G}_L(D)$~\eqref{eq:nmf_scan_local},
the oracle GLRT~\eqref{eq:oracle_theta_true_trueSigma},
and \emph{DopplerGLRTNet} (Doppler regression~\eqref{eq:theta_hat_param} followed by a single normalized correlation~\eqref{eq:theta_score_core}).
All practical detectors use the same global SCM whitening $\widehat{\boldsymbol{\Sigma}}$ estimated on $H_0$ training data, whereas the oracle uses the true covariance $\boldsymbol{\Sigma}$ and the true Doppler $\theta_0$.

Across all disturbance settings in Fig.~\ref{fig:detection_scenarios}, \emph{DopplerGLRTNet} consistently follows the local NMF scan while improving over the single-bin on-grid NMF, with a single correlation per test vector and empirical $H_0$ calibration at the target $P_{\mathrm{fa}}$.

\subsection{Gaussian Clutter: Off-Grid Saturation}
In complex Gaussian clutter under matched distributional assumptions and global SCM whitening estimated from $H_0$ data, the single-bin on-grid NMF saturates under off-grid mismatch ($\theta_0\sim\mathcal{U}(D)$), whereas the dense local scan tightly approximates the off-grid GLRT over the cell.
\emph{DopplerGLRTNet} mitigates this saturation by predicting a continuous Doppler within $D$ and evaluating a single matched filter near the maximizer, thereby approaching scan performance at a fraction of its test-time cost.


\subsection{Compound-Gaussian Clutter}
We introduce texture modulation to obtain compound-Gaussian clutter~\cite{conte1995compoundgaussian,tyler1987scatter}.
At the same $P_{\mathrm{fa}}$, \emph{DopplerGLRTNet} consistently outperforms on-grid NMF and closely tracks the dense local scan, indicating that amortized continuous parameter selection remains effective in heavy-tailed environments.

\section{Conclusion}
\label{sec:conclusion}
We introduced \emph{DopplerGLRTNet}, a neural off-grid GLRT for Doppler off-grid radar target detection.
The proposed detector amortizes the continuous-parameter maximization underlying the off-grid GLRT by learning to predict the off-grid Doppler parameter within a resolution cell and evaluating a \emph{single} normalized matched-filter statistic at the predicted parameter.

Under Gaussian and compound-Gaussian clutter, \emph{DopplerGLRTNet} matches the dense local scan with a single correlation per test vector, while consistently outperforming on-grid NMF at the same empirically calibrated $P_{\mathrm{fa}}$.
Future work will extend the approach to multi-cell detection (full Doppler axis), range-Doppler patches, joint estimation of nuisance parameters, and theoretical guarantees for amortized maximization under model mismatch.

\bibliographystyle{IEEEbib}
\bibliography{refs}

@book{scharf_statistical,
  title     = {Statistical Signal Processing: Detection, Estimation, and Time Series Analysis},
  author    = {Scharf, L. L.},
  year      = {1991},
  publisher = {Addison-Wesley},
  address   = {Reading, MA}
}

@book{kay1998_detection,
  title     = {Fundamentals of Statistical Signal Processing, Volume II: Detection Theory},
  author    = {Kay, S. M.},
  year      = {1998},
  publisher = {Prentice Hall},
  address   = {Upper Saddle River, NJ}
}

@article{kelly1986adaptive,
  author  = {Kelly, E. J.},
  title   = {An Adaptive Detection Algorithm},
  journal = {IEEE Transactions on Aerospace and Electronic Systems},
  year    = {1986},
  volume  = {AES-22},
  number  = {2},
  pages   = {115--127},
  doi     = {10.1109/TAES.1986.310745}
}

@article{robey1992cfar_amf,
  author  = {Robey, F. C. and Fuhrmann, D. R. and Kelly, E. J. and Nitzberg, R.},
  title   = {A {CFAR} Adaptive Matched Filter Detector},
  journal = {IEEE Transactions on Aerospace and Electronic Systems},
  year    = {1992},
  volume  = {28},
  number  = {1},
  pages   = {208--216},
  doi     = {10.1109/7.135446}
}

@article{conte1995compoundgaussian,
  author  = {Conte, E. and Lops, M. and Ricci, G.},
  title   = {Asymptotically Optimum Radar Detection in Compound-{Gaussian} Clutter},
  journal = {IEEE Transactions on Aerospace and Electronic Systems},
  year    = {1995},
  volume  = {31},
  number  = {2},
  pages   = {617--625},
  doi     = {10.1109/7.381910}
}

@article{tyler1987scatter,
  author  = {Tyler, D. E.},
  title   = {A Distribution-Free ${M}$-Estimator of Multivariate Scatter},
  journal = {The Annals of Statistics},
  year    = {1987},
  volume  = {15},
  number  = {1},
  pages   = {234--251},
  doi     = {10.1214/aos/1176350364}
}

@article{develter2024offgridNMF,
  author  = {Develter, P. and Bosse, J. and Rabaste, O. and Forster, P. and Ovarlez, J.-P.},
  journal = {IEEE Transactions on Signal Processing},
  title   = {On the False Alarm Probability of the {N}ormalized {M}atched {F}ilter for Off-Grid Targets: A Geometrical Approach and Its Validity Conditions},
  year    = {2024},
  volume  = {72},
  pages   = {982--996},
  doi     = {10.1109/TSP.2024.3358621}
}

@article{davies1987nuisance,
  author  = {Davies, R. B.},
  title   = {Hypothesis Testing When a Nuisance Parameter Is Present Only Under the Alternative},
  journal = {Biometrika},
  year    = {1987},
  volume  = {74},
  number  = {1},
  pages   = {33--43}
}

@book{adler2007randomfields,
  title     = {Random Fields and Geometry},
  author    = {Adler, R. J. and Taylor, J. E.},
  year      = {2007},
  publisher = {Springer},
  address   = {New York},
  doi       = {10.1007/978-0-387-48116-6}
}

@article{monga2021unrolling,
  author  = {Monga, V. and Li, Y. and Eldar, Y. C.},
  title   = {Algorithm Unrolling: Interpretable, Efficient Deep Learning for Signal and Image Processing},
  journal = {IEEE Signal Processing Magazine},
  year    = {2021},
  volume  = {38},
  number  = {2},
  pages   = {18--44},
  doi     = {10.1109/MSP.2020.3037238}
}

@inproceedings{gregor2010lista,
  title     = {Learning Fast Approximations of Sparse Coding},
  author    = {Gregor, K. and Le Cun, Y.},
  booktitle = {Proceedings of the 27th International Conference on Machine Learning (ICML)},
  year      = {2010},
  pages     = {399--406}
}

@inproceedings{ravanelli2018sincnet,
  title     = {Speaker Recognition from Raw Waveform with {SincNet}},
  author    = {Ravanelli, M. and Bengio, Y.},
  booktitle = {IEEE Spoken Language Technology Workshop (SLT)},
  year      = {2018},
  note      = {arXiv:1808.00158}
}

@article{diskin2024cfarnet,
 author={ Diskin, I. and  Beer, V. and  Okun, U. and Wiesel, A.},
  title   = {{CFARNet}: Deep Learning for Target Detection with Constant False Alarm Rate},
  volume = {223},
   pages = {109543},
issn = {0165-1684},
  journal = {Signal Processing},
  year    = {2024},
  doi     = {10.1016/j.sigpro.2024.109543},
  note    = {}
}

@INPROCEEDINGS{VAEalexis,
  author={Rouzoumka, Y. A. and Terreaux, E. and Morisseau, C. and Ovarlez, J.-P. and Ren, C.},
  booktitle={2025 IEEE International Conference on Acoustics, Speech and Signal Processing}, 
  title={Out-of-Distribution Radar Detection in Compound Clutter and Thermal Noise through Variational Autoencoders}, 
  year={2025},
  volume={},
  number={},
  pages={1-5},
  doi={10.1109/ICASSP49660.2025.10889884}}

@inproceedings{rouzoumka2025complex,
  title     = {Complex-Valued Variational Autoencoders for Radar Detection in Joint Compound {G}aussian Clutter and Thermal Noise},
  author    = {Rouzoumka, Y. A. and Terreaux, E. and Morisseau, C. and Ovarlez, J.-P. and Ren, C.},
  booktitle = {33rd European Signal Processing Conference (EUSIPCO)},
  year      = {2025},
  pages     = {2512--2516}
}

@book{huber2009robust,
  title={Robust Statistics},
  author={Huber, P. J. and Ronchetti, E. M.},
  publisher={Wiley},
  year={2009},
  edition={2},
  isbn={9780470129906}
}

@article{Trottier2025OffGridRD,
  title={Off-Grid Radar Detection Strategy for the Normalized Matched Filter: Achieving the {GLRT} Performance},
  author={ Trottier, S. and  Bosse, J. and  Rabaste, O. and  Forster, P. and Ovarlez, J.-P.},
  journal={33rd European Signal Processing Conference (EUSIPCO)},
  year={2025},
  pages={2557-2561},
  url={https://api.semanticscholar.org/CorpusID:283098358}
}

@book{de2016modern,
	Editor = {Greco, M. S.  and De Maio, A.},
	Month = {Jan},
	Publisher = {SciTech Publishing},
	Title = {Modern Radar Detection Theory},
	Year = {2016}}

\end{document}